\documentclass[11pt]{article}
\usepackage{physpaper}

\begin{document}

\PaperTitle{Charged regular AdS$_\mathbf{3}$ black holes in a scalar-vector-tensor theory\\ coupled to Born-Infeld electrodynamics}
\PaperAuthors{G{\"o}khan Alka\c{c}}
\PaperAffils{%
  \affil{}{Department of Aerospace Engineering, Faculty of Engineering,\\ At{\i}l{\i}m University, 06836 Ankara, T\"{u}rkiye}
}
\PaperEmails{\mail{gokhan.alkac@atilim.edu.tr}}
\PaperAbstract{We construct exact, static, circularly symmetric charged regular AdS$_3$ black hole solutions in a three-dimensional scalar-vector-tensor theory obtained from regularized Gauss-Bonnet couplings. While the Maxwell-charged counterpart of this theory suffers from a logarithmic curvature singularity at the origin, we demonstrate that coupling the action to non-linear Born-Infeld electrodynamics cures this pathology. We obtain the metric function and show that all curvature invariants are finite at the origin. Depending on the value of the hair parameter, the center is smooth, or it has a conical or a BTZ-like singularity with finite curvature. We also discuss the behaviour of the matter fields, the conserved electric charge and the asymptotic form of the solution. For different choices of the parameters, the solution describes horizonless regular geometries, extremal black holes, and black holes with one, two or three horizons.}
\MakeTitle

\section{Introduction}
Despite its remarkable successes, general relativity (GR) still leaves several fundamental questions in gravitational physics unresolved. For instance, while it is well established that black holes behave as thermodynamic systems satisfying the four laws of black hole mechanics \cite{Bardeen:1973gs}, a complete microscopic derivation of their entropy via the counting of underlying quantum states remains an open problem. Another issue directly relevant to the present work is the occurrence of spacetime singularities at the core of black holes. Under the assumptions of the singularity theorems of GR \cite{Penrose:1964wq, Hawking:1970zqf}, namely suitable energy conditions and the existence of trapped surfaces, such singularities are inevitable. Because they lead to pathological divergences in physical observables, it is widely expected that quantum gravitational corrections, or appropriate effective extensions of GR, must resolve them in a UV-complete theory.

A popular way to study these problems in a simpler setup is to lower the spacetime dimensions to three. This approach became very active after the discovery of the Bañados–Teitelboim–Zanelli (BTZ) black hole solution \cite{Banados:1992wn,Banados1993}. The BTZ metric is a solution of three-dimensional (3d) general relativity with a negative cosmological constant. It has all the expected physical properties of a black hole and satisfies the laws of thermodynamics. More importantly, as shown by Strominger \cite{Strominger:1997eq}, one can give a microscopic explanation of its entropy using the AdS$_3$/CFT$_2$ correspondence. The singularity problem can also be addressed in 3d. For example, regular (singularity-free) black hole solutions have been found by coupling gravity to non-linear electrodynamics \cite{Cataldo:2000ns, He:2017ujy, HabibMazharimousavi:2011gh} or a perfect fluid \cite{Estrada:2020tbz}, as well as in certain modified gravity theories \cite{Bueno2021, Bueno:2025dqk}.

In the past few years, it was discovered that the Gauss-Bonnet invariant, $\mathcal{G} = R^2 - 4 R_{\mu\nu}^2+R_{\mu\nu\rho\sigma}^2$, which contributes to field equations only in dimensions $d>4$ when used in an action, can be regularized. Through this regularization, one obtains well-defined scalar-tensor or vector-tensor couplings in 3d and 4d. Remarkably, adding these couplings to the Einstein-Hilbert action leads to exact black hole solutions. For the explicit construction of these theories and their main properties, we refer the reader to \cite{Fernandes:2020nbq, Lu:2020iav, Kobayashi:2020wqy, Hennigar:2020lsl, Ma:2020ufk, Hennigar:2020fkv, Hennigar:2020drx, Alkac:2022zda, Charmousis:2025jpx, Eichhorn:2025pgy, Alkac:2025zzi, Liu:2025dqg, Alkac:2025jhx, Lutfuoglu:2025qkt, Konoplya:2025bte, Charmousis:2026dbi}.

In this paper, we fill a gap in the literature. It was shown in \cite{Alkac:2025zzi} that when the scalar-tensor and vector-tensor couplings are introduced in the action with equal coupling constants, one obtains a static, circularly symmetric regular black hole solution. However, the electrically charged version of this solution, obtained by coupling the theory to Maxwell electrodynamics, remains singular at the center. In \cite{Alkac:2025jhx}, a similar problem was encountered for the regular black holes of a theory with two vector fields, and it was solved by coupling the theory to non-linear Born-Infeld electrodynamics. In this work, we show that the same mechanism works for the scalar-vector-tensor theory of \cite{Alkac:2025zzi}, leading to exact charged AdS$_3$ black hole solutions with finite curvature invariants everywhere. Following the classification of \cite{Bueno2021, Bueno:2025dqk}, depending on the value of the hair parameter, these are regular black holes, or black holes with a conical or a BTZ-like singularity at the origin. Since the solution is asymptotically AdS$_3$, it also provides a natural setup for the study of the thermodynamics of regular charged black holes, including their microscopic entropy along the lines of \cite{Strominger:1997eq, Correa:2010hf, Correa:2011dt, Correa:2012rc, Bravo-Gaete:2015iwa}. As we will discuss in detail, this requires the computation of the conserved charges of the solution, which is a non-trivial problem in the presence of the scalar and vector fields, and we leave it to a separate publication.

The rest of the paper is organized as follows. In Section~\ref{sec:SVT}, we introduce the scalar-vector-tensor theory under consideration and briefly review its charged $\text{AdS}_3$ black hole solution coupled to Maxwell electrodynamics. In Section~\ref{sec:BI}, we couple the theory to non-linear Born-Infeld electrodynamics, derive the exact regular charged $\text{AdS}_3$ black hole solution, and analyze its key physical properties. Finally, in Section~\ref{sec:conc}, we summarize our main conclusions and outline possible directions for future work.

\section{Scalar-vector-tensor theory and its charged solution}\label{sec:SVT}
As mentioned in the introduction, the Gauss-Bonnet invariant, which is given by
\begin{equation}
	\mathcal{G} = R^2 - 4 R_{\mu\nu}^2+R_{\mu\nu\rho\sigma}^2,
\end{equation}
can be regularized to obtain well-defined scalar-tensor and vector-tensor interactions in lower-dimensions ($d<5$). In 3d, they read \cite{Hennigar:2020fkv, Alkac:2022fuc, Alkac:2025zzi}
\begin{equation}
	\begin{aligned}
		\mathcal{G}_{d\to3}^{\text{st}} &= 4 G^{\mu \nu} \varphi_\mu \varphi_\nu-4 X \square \varphi+2 X^2,\\
		\mathcal{G}_{d\to 3}^\text{vt} &=-4 G^{\mu\nu}W_\mu W_\nu-4 W^2 \nabla_\mu W^\mu-2 W^4,
	\end{aligned}\label{scaandvec}
\end{equation}
where $\varphi$ and $W_\mu$ are the scalar and the vector fields that arise due to the regularization respectively. We use the following definitions for a more compact notation
\begin{equation}
	\varphi_\mu \equiv \partial_\mu \varphi, \quad \varphi_{\mu \nu} \equiv \nabla_\mu \nabla_\nu\varphi , \quad X \equiv \partial_\mu \varphi \partial^\mu \varphi, \quad W^2 \equiv W_\mu W^\mu, \quad W^4 \equiv(W^2)^2.
\end{equation}

One can use these interactions to obtain generalizations of the BTZ black hole. With this aim, we consider an action which is given by
\begin{equation}\label{action}
	I=\frac{1}{16 \pi G} \int \dd^3x\sqrt{-g} \left (R-2\Lambda_0
	+\alpha\,\mathcal{G}_{d\to 3}^{\text{st}}
	+\beta\,\mathcal{G}_{d\to 3}^{\text{vt}}\right ),
\end{equation}
where $\Lambda_0$ is the bare cosmological constant, $\alpha$ and $\beta$ are the coupling constants of scalar-tensor and vector-tensor interactions respectively. As the Gauss-Bonnet invariant, the regularized couplings in Eq.~\eqref{scaandvec} lead to second-order field equations. Therefore, the action \eqref{action} belongs to the general class of scalar-vector-tensor theories with second-order field equations constructed in \cite{Heisenberg:2018acv}. We study solutions of the following form
\begin{equation}
	\begin{aligned}
		\dd s^2 &= -N^2(r)f(r) \dd t^2+\frac{\dd r^2}{f(r)}+r^2\dd\theta^2,\\
		W &= w_0(r)\dd t+w_1(r)\dd r, \qquad \varphi = \varphi(r).
	\end{aligned}\label{ansatz}
\end{equation}
We assume that the spacetime is static and circularly symmetric, and matter fields depend only on the radial coordinate $r$. We also take the $\theta$-component of the vector field zero [$W_\theta = w_2 =  0$].

As will be explained in detail in the next section, one can insert the ansatz in \eqref{ansatz} into the action \eqref{action} and obtain a 1d reduced action $I_{\text{red}}= \int \dd{r} L_{\text{red}}$ describing the dynamics of the unknown functions $\Phi_A=\left\{ N, f, w_0, w_1, \varphi \right\}$. The Euler-Lagrange equations following from this reduced action $\fdv{I_\text{red}}{\Phi_A} = 0$ can then be solved with the help of further simplifying assumptions. In \cite{Alkac:2025zzi}, both $\alpha \neq \beta$ and $\alpha = \beta$ cases were studied. In that work, a Noether charge arising from a global symmetry of the reduced action was used to decouple the field equations. However, it was realized later in \cite{Liu:2025dqg} that assuming directly $N=1$ and $W^2=0$ is more practical and leads to the same solution up to redefinitions of the integration constants. Therefore, we adopt this approach here. When the coupling constants are the same ($\alpha = \beta$), assuming $N=1$ and $W^2=0$, one obtains the following solution
\begin{equation}
	f = \frac{(-m-\Lambda_0 r^2)r^2+4 \beta q^2}{r^2+4\beta q}, \qquad w_0 =\frac{2q-f}{2r}, \qquad w_1=\frac{2q-f}{2rf}, \qquad \varphi=\log \left(r / r_0\right).\label{alphabeta}
\end{equation}
Here, $m$ is an integration constant which is related to the mass $M$ of the black hole, and $q$ is the hair parameter resulting from the vector coupling. The integration constant appearing in the equation of the vector field is chosen as $2q$ so that the metric function satisfies $f(0)=q$. As in \cite{Alkac:2025zzi, Alkac:2025jhx}, the exact relation between $m$ and $M$ requires an analysis of the asymptotic charges of the theory, which we comment on in Section \ref{sec:conc}.

The curvature singularities at the center can be studied by checking the behaviour of curvature invariants as $r \to 0$. For the metric in \eqref{ansatz} with $N=1$, the non-vanishing components of the Ricci tensor with mixed indices are
\begin{equation}\label{Ricci}
	R^t_{\ t}=R^r_{\ r}=-\frac{1}{2}\left(f''+\frac{f'}{r}\right), \qquad R^\theta_{\ \theta}=-\frac{f'}{r},
\end{equation}
where primes denote derivatives with respect to $r$. In 3d, the Weyl tensor vanishes identically and the Riemann tensor is completely determined by the Ricci tensor. Furthermore, all curvature invariants can be expressed in terms of $R$, $R^{\mu}_{\ \nu} R^{\nu}_{\ \mu}$ and $R^{\mu}_{\ \nu} R^{\nu}_{\ \rho} R^{\rho}_{\ \mu}$ \cite{Paulos:2010ke, Gurses:2011fv, Bueno:2022lhf}. Since the latter are polynomials in the components \eqref{Ricci}, the finiteness of $f''$ and $f'/r$ at the origin guarantees the finiteness of all curvature invariants. For the solution \eqref{alphabeta}, we have
\begin{equation}
	f=q-\frac{m+q}{4\beta q} r^2+\mathcal{O}\left(r^4\right),
\end{equation}
which gives, for example,
\begin{equation}
	R=\frac{3(m+q)}{2 \beta q}+\mathcal{O}\left(r^2\right),
\end{equation}
and shows that the curvature invariants are finite at the origin. As we will discuss in the next section, the solution is regular at the origin for $q=1$, while it has a conical or a BTZ-like singularity with finite curvature for other values of $q$. To avoid a singularity at a non-zero $r$ value due to a pole of the metric function, one should have $\beta q >0$.

A charged version of this regular black hole solution can be obtained by coupling to Maxwell electrodynamics with the Lagrangian
\begin{equation}\label{Max}
	\mathcal{L}_\text{M} = -\frac{1}{4}F_{\mu\nu} F^{\mu\nu}, \qquad F_{\mu\nu} = 2 \partial_{[\mu} A_{\nu]},
\end{equation}
and taking the gauge field $A$ as
\begin{equation}\label{gauge}
	A = \phi(r) \dd{t},
\end{equation}
where $\phi(r)$ is the electric potential. After obtaining the reduced action with the Maxwell contribution and finding the corresponding Euler-Lagrange equation, the electric potential can be solved directly since it is decoupled from the other unknown functions. The result is the Coulomb potential given by
\begin{equation}\label{coulomb}
	\phi = -q_e \log (r/r_0),\qquad r_0: \text{constant},
\end{equation}
where $q_e$ is the electric charge parameter. As shown in \cite{Alkac:2025zzi} [with the normalization in Eq.~\eqref{Max}], the components of the vector $W$ in terms of the metric function $f$ and the scalar field $\varphi$ remain the same but the metric function now reads
\begin{equation}\label{metmax}
	f=\frac{(-m-\Lambda_0 r^2)r^2+4 \beta q^2-\frac{q_e^2}{2} r^2\log(r/r_0)}{r^2+4\beta q}.
\end{equation}
This is a charged black hole solution but the Ricci scalar now behaves near zero as
\begin{equation}
	R = \frac{3(m + q) + \frac{q_e^2}{4} \left[5 + 6 \log(r/r_0)\right]}{2\beta q} + \mathcal{O}(r^2 \log r).
\end{equation}
As we see, due to the logarithmic form of the electric potential, the solution becomes singular. In the next section, we will couple the theory to Born-Infeld electrodynamics, and remedy this issue.

\section{Coupling to Born-Infeld Electrodynamics and Regular Charged Solution}
\label{sec:BI}

We are now prepared to construct a regular, charged $\text{AdS}_3$ black hole solution. We consider the action
\begin{equation}
	\label{action2}
	I = \frac{1}{16 \pi G} \int \dd^3 x \sqrt{-g} \left( R - 2\Lambda_0 + \beta\,\mathcal{G}_{d\to 3}^{\text{st}} + \beta\,\mathcal{G}_{d\to 3}^{\text{vt}} + \mathcal{L}_{\text{BI}} \right).
\end{equation}
We introduce the scalar-tensor and vector-tensor couplings in Eq.~\eqref{scaandvec} with equal coupling constants, which, as shown in \cite{Alkac:2025zzi}, is required for the regularity of the uncharged solution. For the electromagnetic coupling, we use the Born-Infeld Lagrangian given by
\begin{equation}
	\label{BI}
	\mathcal{L}_{\text{BI}} = b^2 \left( 1 - \sqrt{1 + \frac{1}{2b^2} F_{\mu\nu} F^{\mu\nu}} \right),
\end{equation}
where $b$ is the Born-Infeld parameter. In the limit $b \to \infty$, the standard Maxwell Lagrangian in Eq.~\eqref{Max} is recovered.

\subsection{Solution}
Inserting the metric ansatz from Eq.~\eqref{ansatz} along with the gauge field configuration \eqref{gauge} into the action \eqref{action2}, one obtains a reduced action $I_{\text{red}} = \int \dd{r} \, \mathcal{L}_{\text{red}}$. The reduced Lagrangian $\mathcal{L}_{\text{red}}$ can be simplified by integrating by parts so that it contains at most first derivatives of the unknown functions $\Phi_A = \{ N, f, w_0, w_1, \varphi, \phi \}$. The final result is:
\begin{equation}\label{Lred}
	\begin{aligned}
		\mathcal{L}_{\text{red}} &= \mathcal{L}_{\text{red}}^{\text{EH}} + \beta \mathcal{L}_{\text{red}}^{[\varphi]} + \beta \mathcal{L}_{\text{red}}^{[W]} + \mathcal{L}_{\text{red}}^{[\phi]}, \\
		\mathcal{L}_{\text{red}}^{\text{EH}} &= N \left( \Lambda_0 r + \frac{f'}{2} \right), \\
		\mathcal{L}_{\text{red}}^{[\varphi]} &= \frac{4}{3} N f^2 {\varphi'}^3 \left( 1 - \frac{3}{4} r \varphi' \right) - 2 N' f^2 {\varphi'}^2 \left( 1 - \frac{2}{3} r \varphi' \right) - N f f' {\varphi'}^2 \left( 1 - \frac{2}{3} r \varphi' \right), \\
		\mathcal{L}_{\text{red}}^{[W]} &= \frac{r w_0^4}{N^3 f^2} - N' \left[ \frac{2 r w_0^2 w_1}{N^2} - 2 f^2 w_1^2 (1 + r w_1) \right] - \frac{w_0^2 \left[ (1 + 2 r w_1) f' + 2 f (w_1 + r w_1^2 + r w_1') \right]}{N f} \\
		&\quad + N \left[ f w_1^2 (1 + 2 r w_1) f' + f^2 w_1^2 (2 w_1 + r w_1^2 + 2 r w_1') \right], \\
		\mathcal{L}_{\text{red}}^{[\phi]} &= \frac{b^2 N r}{2} \left( -1 + \sqrt{1 - \frac{{\phi'}^2}{b^2 N^2}} \right).
	\end{aligned}
\end{equation}
Here, $\mathcal{L}_{\text{red}}^{\text{EH}}$ is the contribution from the Einstein-Hilbert term with the cosmological constant, while $\mathcal{L}_{\text{red}}^{[\varphi]}$ and $\mathcal{L}_{\text{red}}^{[W]}$ are the contributions from the scalar-tensor and vector-tensor couplings in Eq.~\eqref{scaandvec}, respectively. The term $\mathcal{L}_{\text{red}}^{[\phi]}$ is the contribution from the Born-Infeld Lagrangian in Eq.~\eqref{BI}.

Since the Euler-Lagrange equations for the unknown functions $\Phi_A = \{ N, f, w_0, w_1, \varphi, \phi \}$, given by $\delta I_{\text{red}} / \delta \Phi_A = 0$, are quite long and cumbersome, we will outline only the main procedure used to obtain the solution below.

Assuming $N=1$, the equation of motion for the Born-Infeld gauge field is
\begin{equation}
	\label{BIequation}
	\phi' + r \phi'' - \frac{1}{b^2} {\phi'}^3 = 0,
\end{equation}
from which the electrostatic scalar potential $\phi(r)$ is found to be \cite{Cataldo:1999wr, Alkac:2022tvc}
\begin{equation}
	\label{BIpotential}
	\phi= -q_e \log \left( \frac{r + \sqrt{r^2 + \frac{q_e^2}{b^2}}}{2 r_0} \right),
\end{equation}
where $q_e$ is the electric charge parameter. Note that the standard Coulomb potential in Eq.~\eqref{coulomb} is recovered in the weak-field limit $b \to \infty$. Now, taking $W^2 = 0$, we find $w_1 = \pm w_0/f$, and we choose the branch
\begin{equation}\label{null}
	w_1 = \frac{w_0}{f}.
\end{equation}
Solving the remaining field equations is still challenging. Therefore, as done in \cite{Alkac:2025zzi}, we assume that the scalar field takes the same form as in the case without vector-tensor coupling:
\begin{equation}
	\varphi=\log \left(r / r_0\right),
\end{equation}
which was first obtained in \cite{Hennigar:2020fkv}. With this choice, the equation of motion for $w_0$ can be integrated, and the components of the vector field are found as
\begin{equation}\label{vector}
	w_0 = \frac{2q-f}{2 r}, \qquad w_1 = \frac{2q-f}{2 r f},
\end{equation}
where $q$ is an integration constant. Note that we use the same profiles for the scalar field $\varphi$ and the vector field components $W_\mu$ (in terms of the metric function) as those used to obtain the uncharged solutions discussed in the previous section [see Eqs.~\eqref{alphabeta}]. With these profiles, all the remaining equations are satisfied identically except the one obtained from the variation with respect to $N$, which is a first-order differential equation for the metric function $f(r)$. The presence of the Born-Infeld potential $\phi(r)$ in Eq.~\eqref{BIpotential} modifies this equation. Integrating it, the exact metric function is obtained as
\begin{equation}
	\label{met}
	f(r) = \frac{r^2 \left[ -\Lambda_0 r^2 - m + \frac{q_e^2}{2} \left( \frac{1}{1+\psi} - \frac{2 \Lambda_0}{b^2} - \operatorname{arctanh}(\psi) \right) + \frac{4 q^2 b^2 \beta}{q_e^2 \psi^2} \right]}{r^2 + 4 q \beta},
\end{equation}
where we have defined $\psi(r) = \sqrt{\frac{b^2 r^2}{q_e^2 + b^2 r^2}}$. This metric function is significantly more complicated than the Maxwell-charged metric function in Eq.~\eqref{metmax}. However, in exchange for this complexity, we obtain a black hole solution with finite curvature invariants everywhere, as we now show.

\subsection{Regularity of the geometry}
Near the origin, the metric function \eqref{met} behaves as
\begin{equation}\label{fexp}
	f(r) = q + f_2 r^2 - \frac{b\, q_e}{4 \beta q} r^3 + \mathcal{O}(r^4), \qquad f_2 = \frac{q_e^2}{8\beta q}-\frac{m+q}{4\beta q}-\frac{\Lambda_0 q_e^2}{4 b^2 \beta q}+\frac{b^2 q}{q_e^2}.
\end{equation}
Therefore, $f''$ and $f'/r$ are finite at the origin, and by the argument given in the previous section, all curvature invariants are finite. For example, the Ricci scalar behaves as
\begin{equation}\label{Rexp}
	R(r) = \frac{3}{4 q} \left[ \frac{2(m+q) - q_e^2}{\beta} - \frac{8 b^2 q^2}{q_e^2} + \frac{2 \Lambda_0 q_e^2}{\beta b^2} \right] + \frac{3 b q_e}{\beta q} r+ \mathcal{O}(r^2),
\end{equation}
and the quadratic and cubic invariants built from \eqref{Ricci} are finite as well. As long as $\beta q > 0$ (which prevents any real root divergence in the denominator at non-zero radial values), the curvature invariants remain finite everywhere. Note that the term linear in $r$ in \eqref{Rexp} originates from the $r^3$ term in \eqref{fexp}, which is absent in the uncharged solution. As a result, the metric is twice continuously differentiable at the origin but not smooth. Here and in the following, regularity refers to the finiteness of the algebraic curvature invariants, which is the standard criterion in the literature \cite{Bueno2021, Bueno:2025dqk}. Differential invariants involving second derivatives of the curvature, such as $\square R$, diverge as $1/r$ at the origin due to the linear term in \eqref{Rexp}, while the tidal forces, which are determined by the Riemann tensor, remain finite. This behaviour can be understood as follows. The Born-Infeld field strength $F_{rt}=\phi'$ is bounded at the origin, $\phi'(0)=-b$, but in 3d the Born-Infeld Lagrangian evaluated on the solution still diverges as $b q_e/r$. In Einstein-Born-Infeld theory, this leads to a curvature singularity with $R \propto 1/r$ \cite{Cataldo:1999wr}. Here, the vector-tensor coupling improves this behaviour such that the curvature stays finite, and the divergence of the source appears only in the $r^3$ term of \eqref{fexp}.

The finiteness of the curvature invariants is necessary for the regularity of a black hole, but the local structure of the center deserves a separate discussion since $f(0)=q$ is not necessarily equal to one. We follow the classification of \cite{Bueno2021, Bueno:2025dqk}, where the same three possibilities arise. For $q>0$, the metric near the origin is approximately $\dd s^2 \simeq -q \dd t^2+\dd r^2/q+r^2\dd\theta^2$. After the rescalings $t \to t/\sqrt{q}$ and $r \to \sqrt{q}\, r$, this becomes the flat metric with an angular coordinate of period $2\pi\sqrt{q}$. Therefore, the center is smooth only for $q=1$, and we call the corresponding solutions regular black holes. For $0<q<1$ and $q>1$, there is a conical singularity at the origin with a deficit and an excess angle respectively, which is the same structure as the one created by a point particle in 3d gravity \cite{Deser:1983tn}. The curvature of a cone is a delta function located at the tip, and therefore, it cannot be detected by the curvature invariants evaluated at $r \neq 0$. For radial geodesics with the conserved energy $E=f\, \dd t/\dd\lambda$, where $\lambda$ is the affine parameter, the geodesic equation reduces to
\begin{equation}
	\left(\frac{\dd r}{\dd \lambda}\right)^2 = E^2 - \kappa f(r),
\end{equation}
where $\kappa=1$ for timelike and $\kappa=0$ for null geodesics. Since $f(0)=q$ is finite, geodesics with $E^2>\kappa q$ reach the origin at a finite value of the affine parameter with $\dd r/\dd \lambda \neq 0$, as in flat space, and the spacetime is geodesically incomplete only at the tip of the cone. For $q<0$, the Killing vector $\partial_t$ is spacelike near the origin, and the surface $r=0$ is a spacelike surface where the circles of constant $r$ shrink to zero size. The curvature is finite, but the local structure is the same as that of the BTZ black hole at $r=0$ \cite{Banados1993}, which is a causal singularity rather than a curvature singularity, and it is called a BTZ-like singularity in \cite{Bueno2021, Bueno:2025dqk}. We will see below that these three cases also correspond to different horizon structures. Note that this distinction is specific to 3d. For a static, spherically symmetric metric in 4d, the angular part of the metric contributes the component $R^{\theta\chi}_{\ \ \theta\chi}=(1-f)/r^2$ to the Riemann tensor, where $(\theta,\chi)$ are the angular coordinates, so that the Kretschmann scalar $R_{\mu\nu\rho\sigma}R^{\mu\nu\rho\sigma}=f''^2+4f'^2/r^2+4(1-f)^2/r^4$ diverges at the origin unless $f(0)=1$. In 3d, there is no such component, and the condition $f(0)=1$ has to be imposed separately.

\subsection{Matter fields and the electric charge}
It is important to distinguish the regularity of the geometry from the regularity of the full field configuration. The scalar field $\varphi = \log(r/r_0)$ diverges logarithmically at the origin, and the scalar kinetic term $X = f/r^2$ diverges as $q/r^2$. The components of the vector field \eqref{vector} behave as $w_0 \simeq q/(2r)$ and $w_1 \simeq 1/(2r)$ near the origin. However, the vector field is null ($W^2=0$) everywhere, and the quantities $\nabla_\mu W^\mu = -f'/(2r)$ and $G^{\mu\nu}W_\mu W_\nu=0$, which are the ones appearing in the action, are finite. At a horizon $r=r_+$, $w_0(r_+)=q/r_+$ is finite while $w_1$ diverges. However, the vector field can be written as $W = w_0 (\dd t + \dd r/f) = w_0\, \dd v$, where $v$ is the ingoing Eddington-Finkelstein coordinate, and therefore, it is regular across the future horizon in the same way as the null one-form $\dd v$. A similar situation occurs for the regular black holes of \cite{Bueno2021}, where the scalar field is linear in the angular coordinate and its kinetic term diverges as $1/r^2$ at the origin. Therefore, the regularity of the $q=1$ solutions should be understood as the regularity of the geometry, and the matter fields that support it have singular components with finite invariants.

The parameter $q_e$ is the conserved electric charge of the Born-Infeld field up to a normalization factor. The equation of motion of the gauge field, $\nabla_\mu \left(F^{\mu\nu}/\sqrt{1+F^2/2b^2}\right)=0$, is a conservation law, and for the potential \eqref{BIpotential} the conserved radial flux is
\begin{equation}
	\frac{r \phi'}{\sqrt{1-\phi'^2/b^2}} = -q_e.
\end{equation}
Defining the electric charge by the Gauss law with the normalization of the action \eqref{action2}, we obtain
\begin{equation}
	Q_e = \frac{1}{16 \pi G} \oint \dd\theta \sqrt{-g}\, \frac{F^{rt}}{\sqrt{1+F^2/2b^2}}=\frac{q_e}{8 G}.
\end{equation}
The same relation holds for the Maxwell case with the normalization in Eq.~\eqref{Max}, which is the reason why we use the Lagrangian $-F^2/4$ for the Maxwell theory, the weak-field limit of \eqref{BI}.

\subsection{Asymptotic behaviour}
At large $r$, we have $\psi \to 1$ and $\operatorname{arctanh}(\psi) \simeq \log (2br/q_e)$. As a result, the metric function \eqref{met} behaves as
\begin{equation}\label{asymp}
	f(r) = -\Lambda_0 r^2 - \frac{q_e^2}{2}\log\left(\frac{2 b r}{q_e}\right) - \left[m - \frac{q_e^2}{4} + \frac{\Lambda_0 q_e^2}{b^2} - \frac{4 b^2 \beta q^2}{q_e^2}-4\Lambda_0 \beta q\right] + \mathcal{O}\left(\frac{\log r}{r^2}\right).
\end{equation}
Therefore, the solution is asymptotically AdS$_3$ with the AdS radius $\ell^2=-1/\Lambda_0$, and the sub-leading terms have the same logarithmic form as those of the charged BTZ black hole \cite{Martinez:1999qi}. The combination in the square brackets plays the role of the mass parameter of the BTZ black hole. However, the identification of the conserved mass requires the relaxed asymptotic conditions of \cite{Martinez:1999qi} due to the logarithmic term, and also the contributions of the scalar and vector fields, which do not decay at infinity [$X = -\Lambda_0 + \mathcal{O}(\log r/r^2)$ and $w_0 = -\Lambda_0 r/2 + \mathcal{O}(\log r/r)$]. We will come back to this point in Section \ref{sec:conc}.

\subsection{Horizon structure}
\begin{figure}[t]
	\centering
	\includegraphics[width=0.74\textwidth]{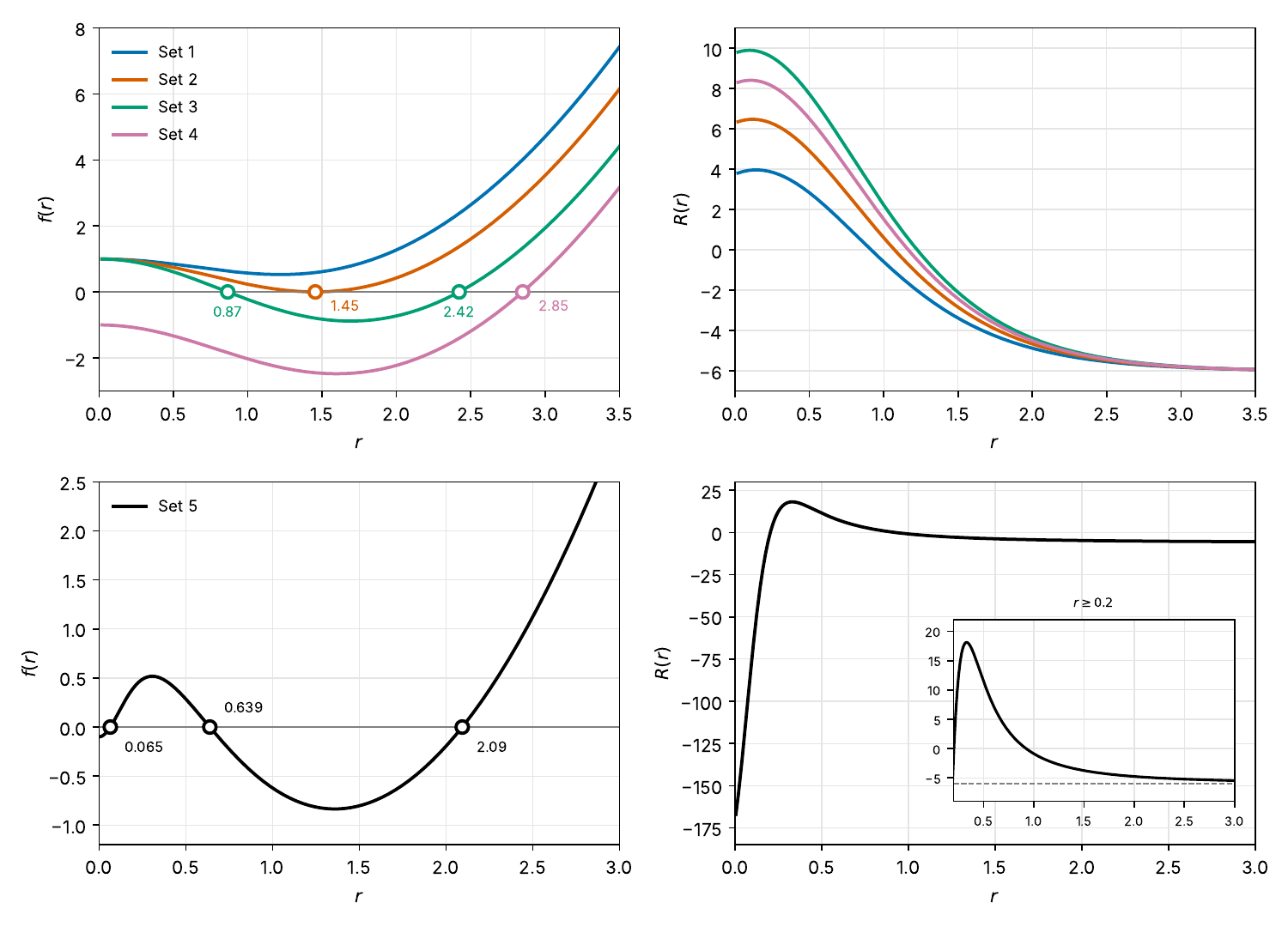}
	\caption{Metric functions $f(r)$ (left panels) and the corresponding Ricci scalars $R(r)$ (right panels) for the five parameter sets defined in Eq.~\eqref{para}. Sets 1, 2, 3, 4 (upper panels) and 5 (lower panels) correspond to a horizonless geometry, an extremal black hole, a black hole with two distinct horizons, a black hole with a single horizon, and a black hole with three horizons, respectively. The zeros of $f(r)$ are marked with circles: $r_\pm = 2.4223,\, 0.8655$ for Set 3, $r_{\text{ext}}=1.4547$ for Set 2, $r_+=2.8496$ for Set 4, and $r \simeq 0.0650,\, 0.6386,\, 2.0941$ for Set 5. All solutions have a finite Ricci scalar at $r = 0$, and $R \to 6\Lambda_0$ at infinity (dashed line in the inset of the lower right panel, which shows the region $r \geq 0.2$ since $R(0) \propto 1/q$ is large for Set 5).}
	\label{fig}
\end{figure}

An analytical study of the horizon structure corresponding to the metric function in Eq.~\eqref{met} is quite involved. However, a general statement follows from the behaviour of $f(r)$ at the origin and at infinity: $f(0)=q$ and $f \to -\Lambda_0 r^2 \to +\infty$ for $\Lambda_0<0$. For $q>0$, the number of horizons is even. In all the cases we have examined, there are either no horizons, two distinct horizons, or a single degenerate horizon (extremal black hole). For $q<0$, which requires $\beta<0$ due to the condition $\beta q>0$, the number of horizons is odd. Black holes with a single horizon exist only in this case, and black holes with three horizons are also possible. We can numerically verify that all these possibilities are realized. Taking $\Lambda_0 = -1$, the following sets of the remaining parameters exemplify them:
\begin{equation}
	\label{para}
	\begin{aligned}
		\text{Set 1:}& \qquad (m = 7, \, q = 1, \, \beta = 1, \, q_e = 1, \, b = 1), \\
		\text{Set 2:}& \qquad (m = m_{\text{ext}} \simeq 8.6958, \, q = 1, \, \beta = 1, \, q_e = 1, \, b = 1), \\
		\text{Set 3:}& \qquad (m = 11, \, q = 1, \, \beta = 1, \, q_e = 1, \, b = 1), \\
		\text{Set 4:}& \qquad (m = 4, \, q = -1, \, \beta = -1, \, q_e = 1, \, b = 1), \\
		\text{Set 5:}& \qquad (m = 4, \, q = -0.1, \, \beta = -0.25, \, q_e = 3, \, b = 2).
	\end{aligned}
\end{equation}
Here, the extremal value $m_{\text{ext}}$ is determined by the conditions $f(r_{\text{ext}})=0$ and $f'(r_{\text{ext}})=0$, which give $r_{\text{ext}} \simeq 1.4547$. For $m<m_{\text{ext}}$, the geometry is horizonless (Set 1), and for $m>m_{\text{ext}}$, there are two distinct horizons (Set 3). Set 4 has a single horizon and Set 5 has three horizons. For a black hole with a conical singularity at the origin, one can take, for example, $q=2$ with the other parameters as in Set 3 and $m=30$: there are two horizons at $r \simeq 1.161$ and $r \simeq 3.528$ (the extremal value is $m_{\text{ext}} \simeq 24.54$ for $q=2$), and the angular coordinate has the period $2\pi\sqrt{2}$ at the origin. The behavior of the metric function $f(r)$ and the corresponding Ricci scalar $R$ for these choices are illustrated in Figure~\ref{fig}, where the horizons and the degenerate horizon of the extremal black hole are marked. A summary of the properties of the solution is given in Table \ref{tab}.
\begin{table}[t]
	\centering
	\small
	\begin{tabular}{lccc}
		\toprule
		Parameter range & Number of horizons & Local structure at $r=0$ & Example \\
		\midrule
		$q=1$, $\beta>0$, $m<m_{\text{ext}}$ & 0 & smooth (regular horizonless geometry) & Set 1 \\
		$q=1$, $\beta>0$, $m=m_{\text{ext}}$ & 1 (degenerate) & smooth (regular extremal black hole) & Set 2 \\
		$q=1$, $\beta>0$, $m>m_{\text{ext}}$ & 2 & smooth (regular black hole) & Set 3 \\
		$0<q\neq1$, $\beta>0$ & 0, 1 (degenerate) or 2 & conical singularity, period $2\pi\sqrt{q}$ & $q=2$ example \\
		$q<0$, $\beta<0$ & 1 or 3 & BTZ-like singularity & Sets 4, 5 \\
		\bottomrule
	\end{tabular}
	\caption{Horizon structure and the local structure of the center of the solution \eqref{met} for $\Lambda_0<0$. In all cases $\beta q>0$ is required and all curvature invariants are finite at $r=0$; the solutions are regular black holes in the sense of \cite{Bueno2021, Bueno:2025dqk} only for $q=1$. For $q>0$, the number of horizons depends on the value of $m$ in the same way as for $q=1$, with a different extremal value $m_{\text{ext}}(q,\beta,q_e,b)$. The number of horizons is even for $q>0$ and odd for $q<0$; the entries for $q>0$ are the possibilities realized in all the cases we have examined.}
	\label{tab}
\end{table}

\section{Conclusions}\label{sec:conc}

In this paper, we have constructed an exact, static, circularly symmetric charged regular $\text{AdS}_3$ black hole solution in a three-dimensional scalar-vector-tensor theory obtained from regularized Gauss-Bonnet couplings with equal coupling constants. While coupling this theory to standard linear Maxwell electrodynamics yields a solution with a logarithmic curvature singularity at the origin, we have shown that this pathology is cured by coupling the theory to non-linear Born-Infeld electrodynamics. We derived the metric function $f(r)$ and demonstrated that all curvature invariants remain finite at the origin. We also discussed the local structure of the center: it is smooth for $q=1$, in which case the solutions are regular black holes, while there is a conical singularity for $0<q\neq1$ and a BTZ-like singularity for $q<0$. In addition, we discussed the behaviour of the matter fields, the conserved electric charge and the asymptotic form of the solution. Furthermore, our analysis revealed that depending on the choice of parameters, the solution describes horizonless geometries, extremal black holes, or black holes with one, two or three horizons.

The natural next step is the study of the thermodynamics of this solution. In this paper, we followed the common practice in the literature \cite{Alkac:2025zzi, Alkac:2025jhx} and introduced $m$ as an integration constant that reduces to the BTZ mass parameter in the appropriate limit. The computation of the conserved mass $M$ and the entropy $S$ is a non-trivial problem for the following reasons. First, the scalar and the vector fields do not decay at infinity, and therefore they contribute to the boundary terms that define the conserved charges. Second, the logarithmic term in the asymptotic expansion \eqref{asymp} requires the relaxed boundary conditions and the renormalization scale of the charged BTZ black hole \cite{Martinez:1999qi}. Third, the vector field is null, $W=w_0 (\dd t+ \dd r/f)$, which is regular on the future horizon but not under the Wick rotation to Euclidean signature, since $w_0$ does not vanish at the horizon. This also prevents a direct application of the Wald formula \cite{Wald:1993nt, Iyer:1994ys}, which assumes that the fields are regular at the bifurcation surface. A method that deals with such difficulties was recently developed in \cite{Bahamonde:2025qtc} for a black hole solution of a Weyl-geometry Gauss-Bonnet theory in five dimensions, where the regularity of the vector field at the horizon is imposed by the choice of the integration constants, and the entropy is obtained from the Euclidean action. A general variational treatment of the infrared divergences of on-shell actions, which does not require a background subtraction, was given in \cite{Arroyo:2026pdu}. The application of these methods to the solution found here, together with the methods of \cite{Alkac:2023mvr, Alkac:2024hvu}, will be presented in an upcoming publication \cite{inprep}. Once the mass is known, the solution being asymptotically AdS$_3$, a microscopic derivation of the entropy along the lines of \cite{Correa:2010hf, Correa:2011dt, Correa:2012rc, Bravo-Gaete:2015iwa}, with the soliton obtained by a double Wick rotation as the ground state, becomes possible. Other natural extensions include constructing rotating versions of this solution and studying its properties under different types of non-linear electrodynamics.

\paragraph*{Acknowledgements} This work was supported by the Turkish Academy of Sciences through the Outstanding Young Scientists Award Programme (TÜBA-GEBİP 2025). During the preparation of this work, the author utilized the AI language model Gemini 2.5 Flash for proofreading, polishing LaTeX syntax, and refining phrasing for clarity and accessibility. During the revision, the AI language model Claude (Anthropic) was utilized for proofreading, \LaTeX{} formatting and the preparation of Figure \ref{fig}. All technical derivations, physical results, and scientific conclusions were performed, verified, and finalized solely by the author.

\bibliographystyle{utphys}
\bibliography{ref}

\providecommand{\href}[2]{#2}\begingroup\raggedright\begin{thebibliography}{10}

\bibitem{Bardeen:1973gs}
J.~M. Bardeen, B.~Carter, and S.~W. Hawking, ``{The Four laws of black hole
  mechanics},'' \href{https://dx.doi.org/10.1007/BF01645742}{{\em Commun. Math.
  Phys.} {\bfseries 31} (1973) 161--170}.

\bibitem{Penrose:1964wq}
R.~Penrose, ``{Gravitational collapse and space-time singularities},''
  \href{https://dx.doi.org/10.1103/PhysRevLett.14.57}{{\em Phys. Rev. Lett.}
  {\bfseries 14} (1965) 57--59}.

\bibitem{Hawking:1970zqf}
S.~W. Hawking and R.~Penrose, ``{The Singularities of gravitational collapse
  and cosmology},'' \href{https://dx.doi.org/10.1098/rspa.1970.0021}{{\em Proc.
  Roy. Soc. Lond. A} {\bfseries 314} (1970) 529--548}.

\bibitem{Banados:1992wn}
M.~Banados, C.~Teitelboim, and J.~Zanelli, ``{The Black hole in
  three-dimensional space-time},''
  \href{https://dx.doi.org/10.1103/PhysRevLett.69.1849}{{\em Phys. Rev. Lett.}
  {\bfseries 69} (1992) 1849--1851},
  \href{https://arxiv.org/abs/hep-th/9204099}{{\ttfamily
  arXiv:hep-th/9204099}}.

\bibitem{Banados1993}
M.~Banados, M.~Henneaux, C.~Teitelboim, and J.~Zanelli, ``{Geometry of the
  (2+1) black hole},'' \href{https://dx.doi.org/10.1103/PhysRevD.48.1506}{{\em
  Phys. Rev. D} {\bfseries 48} (1993) 1506--1525},
  \href{https://arxiv.org/abs/gr-qc/9302012}{{\ttfamily arXiv:gr-qc/9302012}}.
  [Erratum: Phys.Rev.D 88, 069902 (2013)].

\bibitem{Strominger:1997eq}
A.~Strominger, ``{Black hole entropy from near horizon microstates},''
  \href{https://dx.doi.org/10.1088/1126-6708/1998/02/009}{{\em JHEP} {\bfseries
  02} (1998) 009}, \href{https://arxiv.org/abs/hep-th/9712251}{{\ttfamily
  arXiv:hep-th/9712251}}.

\bibitem{Cataldo:2000ns}
M.~Cataldo and A.~Garcia, ``{Regular (2+1)-dimensional black holes within
  nonlinear electrodynamics},''
  \href{https://dx.doi.org/10.1103/PhysRevD.61.084003}{{\em Phys. Rev. D}
  {\bfseries 61} (2000) 084003},
  \href{https://arxiv.org/abs/hep-th/0004177}{{\ttfamily
  arXiv:hep-th/0004177}}.

\bibitem{He:2017ujy}
Y.~He and M.-S. Ma, ``{$(2+1)$-dimensional regular black holes with nonlinear
  electrodynamics sources},''
  \href{https://dx.doi.org/10.1016/j.physletb.2017.09.044}{{\em Phys. Lett. B}
  {\bfseries 774} (2017) 229--234},
  \href{https://arxiv.org/abs/1709.09473}{{\ttfamily arXiv:1709.09473
  [gr-qc]}}.

\bibitem{HabibMazharimousavi:2011gh}
S.~Habib~Mazharimousavi, M.~Halilsoy, and T.~Tahamtan, ``{Regular charged black
  hole construction in 2+1 -dimensions},''
  \href{https://dx.doi.org/10.1016/j.physleta.2012.01.001}{{\em Phys. Lett. A}
  {\bfseries 376} (2012) 893--898},
  \href{https://arxiv.org/abs/1107.0242}{{\ttfamily arXiv:1107.0242 [gr-qc]}}.

\bibitem{Estrada:2020tbz}
M.~Estrada and F.~Tello-Ortiz, ``{A new model of regular black hole in (2+1)
  dimensions},'' \href{https://dx.doi.org/10.1209/0295-5075/ac0ed0}{{\em EPL}
  {\bfseries 135} no.~2, (2021) 20001},
  \href{https://arxiv.org/abs/2012.05068}{{\ttfamily arXiv:2012.05068
  [gr-qc]}}.

\bibitem{Bueno2021}
P.~Bueno, P.~A. Cano, J.~Moreno, and G.~van~der Velde, ``{Regular black holes
  in three dimensions},''
  \href{https://dx.doi.org/10.1103/PhysRevD.104.L021501}{{\em Phys. Rev. D}
  {\bfseries 104} no.~2, (2021) L021501},
  \href{https://arxiv.org/abs/2104.10172}{{\ttfamily arXiv:2104.10172
  [gr-qc]}}.

\bibitem{Bueno:2025dqk}
P.~Bueno, O.~Lasso~Andino, J.~Moreno, and G.~van~der Velde, ``{On regular
  charged black holes in three dimensions},''
  \href{https://dx.doi.org/10.1007/JHEP08(2025)132}{{\em JHEP} {\bfseries 08}
  (2025) 132}, \href{https://arxiv.org/abs/2503.02930}{{\ttfamily
  arXiv:2503.02930 [gr-qc]}}.

\bibitem{Fernandes:2020nbq}
P.~G.~S. Fernandes, P.~Carrilho, T.~Clifton, and D.~J. Mulryne, ``{Derivation
  of regularized field equations for the einstein-gauss-bonnet theory in four
  dimensions},'' \href{https://dx.doi.org/10.1103/PhysRevD.102.024025}{{\em
  Phys. Rev. D} {\bfseries 102} no.~2, (2020) 024025},
  \href{https://arxiv.org/abs/2004.08362}{{\ttfamily arXiv:2004.08362
  [gr-qc]}}.

\bibitem{Lu:2020iav}
H.~Lu and Y.~Pang, ``{Horndeski gravity as $D \to 4$ limit of Gauss-Bonnet},''
  \href{https://dx.doi.org/10.1016/j.physletb.2020.135717}{{\em Phys. Lett. B}
  {\bfseries 809} (2020) 135717},
  \href{https://arxiv.org/abs/2003.11552}{{\ttfamily arXiv:2003.11552
  [gr-qc]}}.

\bibitem{Kobayashi:2020wqy}
T.~Kobayashi, ``{Effective scalar-tensor description of regularized lovelock
  gravity in four dimensions},''
  \href{https://dx.doi.org/10.1088/1475-7516/2020/07/013}{{\em JCAP} {\bfseries
  07} (2020) 013}, \href{https://arxiv.org/abs/2003.12771}{{\ttfamily
  arXiv:2003.12771 [gr-qc]}}.

\bibitem{Hennigar:2020lsl}
R.~A. Hennigar, D.~Kubiz\v{n}\'ak, R.~B. Mann, and C.~Pollack, ``{On taking the
  $D \to 4$ limit of Gauss-Bonnet gravity: theory and solutions},''
  \href{https://dx.doi.org/10.1007/JHEP07(2020)027}{{\em JHEP} {\bfseries 07}
  (2020) 027}, \href{https://arxiv.org/abs/2004.09472}{{\ttfamily
  arXiv:2004.09472 [gr-qc]}}.

\bibitem{Ma:2020ufk}
L.~Ma and H.~Lu, ``{Vacua and exact solutions in lower-d limits of egb},''
  \href{https://dx.doi.org/10.1140/epjc/s10052-020-08780-4}{{\em Eur. Phys. J.
  C} {\bfseries 80} no.~12, (2020) 1209},
  \href{https://arxiv.org/abs/2004.14738}{{\ttfamily arXiv:2004.14738
  [gr-qc]}}.

\bibitem{Hennigar:2020fkv}
R.~A. Hennigar, D.~Kubiznak, R.~B. Mann, and C.~Pollack, ``{Lower-dimensional
  Gauss{\textendash}Bonnet gravity and BTZ black holes},''
  \href{https://dx.doi.org/10.1016/j.physletb.2020.135657}{{\em Phys. Lett. B}
  {\bfseries 808} (2020) 135657},
  \href{https://arxiv.org/abs/2004.12995}{{\ttfamily arXiv:2004.12995
  [gr-qc]}}.

\bibitem{Hennigar:2020drx}
R.~A. Hennigar, D.~Kubiznak, and R.~B. Mann, ``{Rotating gauss-bonnet btz black
  holes},'' \href{https://dx.doi.org/10.1088/1361-6382/abce48}{{\em Class.
  Quant. Grav.} {\bfseries 38} no.~3, (2021) 03LT01},
  \href{https://arxiv.org/abs/2005.13732}{{\ttfamily arXiv:2005.13732
  [gr-qc]}}.

\bibitem{Alkac:2022zda}
G.~Alkac and G.~Suer, ``{3D Lovelock gravity and the holographic c-theorem},''
  \href{https://dx.doi.org/10.1103/PhysRevD.107.046014}{{\em Phys. Rev. D}
  {\bfseries 107} no.~4, (2023) 046014},
  \href{https://arxiv.org/abs/2211.12450}{{\ttfamily arXiv:2211.12450
  [hep-th]}}.

\bibitem{Charmousis:2025jpx}
C.~Charmousis, P.~G.~S. Fernandes, and M.~Hassaine, ``{Proca theory of
  four-dimensional regularized Gauss-Bonnet gravity and black holes with
  primary hair},'' \href{https://dx.doi.org/10.1103/9f2w-3kly}{{\em Phys. Rev.
  D} {\bfseries 111} no.~12, (2025) 124008},
  \href{https://arxiv.org/abs/2504.13084}{{\ttfamily arXiv:2504.13084
  [gr-qc]}}.

\bibitem{Eichhorn:2025pgy}
A.~Eichhorn and P.~G.~S. Fernandes, ``{Regular black holes without
  mass-inflation instability and gravastars from modified gravity},''
  \href{https://dx.doi.org/10.1103/nqz2-88zf}{{\em Phys. Rev. D} {\bfseries
  113} no.~8, (2026) L081501},
  \href{https://arxiv.org/abs/2508.00686}{{\ttfamily arXiv:2508.00686
  [gr-qc]}}.

\bibitem{Alkac:2025zzi}
G.~Alkac, M.~Mesta, and G.~Unal, ``{AdS$_3$ black holes with primary Proca hair
  from regularized Gauss-Bonnet coupling},''
  \href{https://dx.doi.org/10.1103/1ws1-x8z3}{{\em Phys. Rev. D} {\bfseries
  112} no.~8, (2025) 084055},
  \href{https://arxiv.org/abs/2508.03386}{{\ttfamily arXiv:2508.03386
  [hep-th]}}.

\bibitem{Liu:2025dqg}
J.-Z. Liu, S.-J. Yang, C.-C. Zhu, and Y.-X. Liu, ``{D-dimensional black holes
  in extended Gauss-Bonnet gravity},''
  \href{https://dx.doi.org/10.1016/j.physletb.2026.140544}{{\em Phys. Lett. B}
  {\bfseries 878} (2026) 140544},
  \href{https://arxiv.org/abs/2508.04292}{{\ttfamily arXiv:2508.04292
  [gr-qc]}}.

\bibitem{Alkac:2025jhx}
G.~Alkac, M.~Mesta, and G.~Unal, ``{Regular AdS3 black holes from a regularized
  Gauss-Bonnet coupling},''
  \href{https://dx.doi.org/10.1016/j.physletb.2025.140107}{{\em Phys. Lett. B}
  {\bfseries 872} (2026) 140107},
  \href{https://arxiv.org/abs/2508.14010}{{\ttfamily arXiv:2508.14010
  [hep-th]}}.

\bibitem{Lutfuoglu:2025qkt}
B.~C. L{\"u}tf{\"u}o{\u{g}}lu, ``{Long-lived quasinormal modes and echoes in
  the Einstein{\textendash}Gauss{\textendash}Bonnet{\textendash}Proca
  theory},'' \href{https://dx.doi.org/10.1140/epjc/s10052-025-14839-x}{{\em
  Eur. Phys. J. C} {\bfseries 85} no.~9, (2025) 1076},
  \href{https://arxiv.org/abs/2508.19194}{{\ttfamily arXiv:2508.19194
  [gr-qc]}}.

\bibitem{Konoplya:2025bte}
R.~A. Konoplya, D.~Ovchinnikov, and J.~Schee, ``{Primary Proca hair and the
  double-peak optics of black holes},''
  \href{https://dx.doi.org/10.1103/7xy2-ymw9}{{\em Phys. Rev. D} {\bfseries
  113} no.~2, (2026) 024059},
  \href{https://arxiv.org/abs/2510.05947}{{\ttfamily arXiv:2510.05947
  [gr-qc]}}.

\bibitem{Charmousis:2026dbi}
C.~Charmousis, P.~G.~S. Fernandes, and M.~Hassaine, ``{Effective cosmological
  constant as black hole primary hair},''
  \href{https://dx.doi.org/10.1103/4sn9-hbhc}{{\em Phys. Rev. D} {\bfseries
  113} no.~12, (2026) 124063},
  \href{https://arxiv.org/abs/2603.25598}{{\ttfamily arXiv:2603.25598
  [gr-qc]}}.

\bibitem{Correa:2010hf}
F.~Correa, C.~Martinez, and R.~Troncoso, ``{Scalar solitons and the microscopic
  entropy of hairy black holes in three dimensions},''
  \href{https://dx.doi.org/10.1007/JHEP01(2011)034}{{\em JHEP} {\bfseries 01}
  (2011) 034}, \href{https://arxiv.org/abs/1010.1259}{{\ttfamily
  arXiv:1010.1259 [hep-th]}}.

\bibitem{Correa:2011dt}
F.~Correa, C.~Martinez, and R.~Troncoso, ``{Hairy Black Hole Entropy and the
  Role of Solitons in Three Dimensions},''
  \href{https://dx.doi.org/10.1007/JHEP02(2012)136}{{\em JHEP} {\bfseries 02}
  (2012) 136}, \href{https://arxiv.org/abs/1112.6198}{{\ttfamily
  arXiv:1112.6198 [hep-th]}}.

\bibitem{Correa:2012rc}
F.~Correa, A.~Fa\'undez, and C.~Mart\'\i{}nez, ``{Rotating hairy black hole and
  its microscopic entropy in three spacetime dimensions},''
  \href{https://dx.doi.org/10.1103/PhysRevD.87.027502}{{\em Phys. Rev. D}
  {\bfseries 87} no.~2, (2013) 027502},
  \href{https://arxiv.org/abs/1211.4878}{{\ttfamily arXiv:1211.4878 [hep-th]}}.

\bibitem{Bravo-Gaete:2015iwa}
M.~Bravo-Gaete, S.~Gomez, and M.~Hassaine, ``{Cardy formula for charged black
  holes with anisotropic scaling},''
  \href{https://dx.doi.org/10.1103/PhysRevD.92.124002}{{\em Phys. Rev. D}
  {\bfseries 92} no.~12, (2015) 124002},
  \href{https://arxiv.org/abs/1510.04084}{{\ttfamily arXiv:1510.04084
  [hep-th]}}.

\bibitem{Alkac:2022fuc}
G.~Alkac, G.~D. Ozen, and G.~Suer, ``{Lower-dimensional limits of cubic
  lovelock gravity},''
  \href{https://dx.doi.org/10.1016/j.nuclphysb.2022.116027}{{\em Nucl. Phys. B}
  {\bfseries 985} (2022) 116027},
  \href{https://arxiv.org/abs/2203.01811}{{\ttfamily arXiv:2203.01811
  [gr-qc]}}.

\bibitem{Heisenberg:2018acv}
L.~Heisenberg, ``{Scalar-Vector-Tensor Gravity Theories},''
  \href{https://dx.doi.org/10.1088/1475-7516/2018/10/054}{{\em JCAP} {\bfseries
  10} (2018) 054}, \href{https://arxiv.org/abs/1801.01523}{{\ttfamily
  arXiv:1801.01523 [gr-qc]}}.

\bibitem{Paulos:2010ke}
M.~F. Paulos, ``{New massive gravity extended with an arbitrary number of
  curvature corrections},''
  \href{https://dx.doi.org/10.1103/PhysRevD.82.084042}{{\em Phys. Rev. D}
  {\bfseries 82} (2010) 084042},
  \href{https://arxiv.org/abs/1005.1646}{{\ttfamily arXiv:1005.1646 [hep-th]}}.

\bibitem{Gurses:2011fv}
M.~Gurses, T.~C. Sisman, and B.~Tekin, ``{Some exact solutions of all
  $f(R_{\mu\nu})$ theories in three dimensions},''
  \href{https://dx.doi.org/10.1103/PhysRevD.86.024001}{{\em Phys. Rev. D}
  {\bfseries 86} (2012) 024001},
  \href{https://arxiv.org/abs/1112.6346}{{\ttfamily arXiv:1112.6346 [hep-th]}}.

\bibitem{Bueno:2022lhf}
P.~Bueno, P.~A. Cano, Q.~Llorens, J.~Moreno, and G.~van~der Velde, ``{Aspects
  of three-dimensional higher curvature gravities},''
  \href{https://dx.doi.org/10.1088/1361-6382/ac6cbf}{{\em Class. Quant. Grav.}
  {\bfseries 39} no.~12, (2022) 125002},
  \href{https://arxiv.org/abs/2201.07266}{{\ttfamily arXiv:2201.07266
  [hep-th]}}.

\bibitem{Cataldo:1999wr}
M.~Cataldo and A.~Garcia, ``{Three dimensional black hole coupled to the
  Born-Infeld electrodynamics},''
  \href{https://dx.doi.org/10.1016/S0370-2693(99)00441-4}{{\em Phys. Lett. B}
  {\bfseries 456} (1999) 28--33},
  \href{https://arxiv.org/abs/hep-th/9903257}{{\ttfamily
  arXiv:hep-th/9903257}}.

\bibitem{Alkac:2022tvc}
G.~Alkac, M.~K. Gumus, and M.~A. Olpak, ``{Generalized black holes in 3D
  Kerr-Schild double copy},''
  \href{https://dx.doi.org/10.1103/PhysRevD.106.026013}{{\em Phys. Rev. D}
  {\bfseries 106} no.~2, (2022) 026013},
  \href{https://arxiv.org/abs/2205.08503}{{\ttfamily arXiv:2205.08503
  [hep-th]}}.

\bibitem{Deser:1983tn}
S.~Deser, R.~Jackiw, and G.~'t~Hooft, ``{Three-Dimensional Einstein Gravity:
  Dynamics of Flat Space},''
  \href{https://dx.doi.org/10.1016/0003-4916(84)90085-X}{{\em Annals Phys.}
  {\bfseries 152} (1984) 220}.

\bibitem{Martinez:1999qi}
C.~Martinez, C.~Teitelboim, and J.~Zanelli, ``{Charged rotating black hole in
  three space-time dimensions},''
  \href{https://dx.doi.org/10.1103/PhysRevD.61.104013}{{\em Phys. Rev. D}
  {\bfseries 61} (2000) 104013},
  \href{https://arxiv.org/abs/hep-th/9912259}{{\ttfamily
  arXiv:hep-th/9912259}}.

\bibitem{Wald:1993nt}
R.~M. Wald, ``{Black hole entropy is the Noether charge},''
  \href{https://dx.doi.org/10.1103/PhysRevD.48.R3427}{{\em Phys. Rev. D}
  {\bfseries 48} no.~8, (1993) R3427--R3431},
  \href{https://arxiv.org/abs/gr-qc/9307038}{{\ttfamily arXiv:gr-qc/9307038}}.

\bibitem{Iyer:1994ys}
V.~Iyer and R.~M. Wald, ``{Some properties of Noether charge and a proposal for
  dynamical black hole entropy},''
  \href{https://dx.doi.org/10.1103/PhysRevD.50.846}{{\em Phys. Rev. D}
  {\bfseries 50} (1994) 846--864},
  \href{https://arxiv.org/abs/gr-qc/9403028}{{\ttfamily arXiv:gr-qc/9403028}}.

\bibitem{Bahamonde:2025qtc}
S.~Bahamonde and M.~Ba{\~n}ados, ``{An exact five dimensional Weyl-geometry
  Gauss-Bonnet black hole},''
  \href{https://dx.doi.org/10.1016/j.physletb.2025.139869}{{\em Phys. Lett. B}
  {\bfseries 869} (2025) 139869},
  \href{https://arxiv.org/abs/2504.02230}{{\ttfamily arXiv:2504.02230
  [gr-qc]}}.

\bibitem{Arroyo:2026pdu}
M.~Arroyo, D.~Arteaga, M.~Ba{\~n}ados, and A.~Faraggi, ``{The variational
  approach to infrared divergencies: Applications to black hole action
  integrals and holographic Wilson loops},''
  \href{https://arxiv.org/abs/2607.14330}{{\ttfamily arXiv:2607.14330
  [hep-th]}}.

\bibitem{Alkac:2023mvr}
G.~Alka{\c{c}}, G.~D. {\"O}zen, H.~{\"O}z{\c{s}}ahin, G.~S{\"u}er, and M.~Tek,
  ``{Scaling symmetry, Smarr relation, and the extended first law in
  lower-dimensional Lovelock gravity},''
  \href{https://dx.doi.org/10.1016/j.nuclphysb.2024.116535}{{\em Nucl. Phys. B}
  {\bfseries 1002} (2024) 116535},
  \href{https://arxiv.org/abs/2308.00620}{{\ttfamily arXiv:2308.00620
  [hep-th]}}.

\bibitem{Alkac:2024hvu}
G.~Alkac, L.~Guajardo, and H.~Ozsahin, ``{Microscopic entropy of static black
  holes in 3D Lovelock gravities},''
  \href{https://dx.doi.org/10.1103/PhysRevD.111.044006}{{\em Phys. Rev. D}
  {\bfseries 111} no.~4, (2025) 044006},
  \href{https://arxiv.org/abs/2409.03865}{{\ttfamily arXiv:2409.03865
  [hep-th]}}.

\bibitem{inprep}
G.~Alkac, M.~Mesta, and G.~Unal.
\newblock in preparation.

\end{thebibliography}\endgroup

\end{document}